# Self-Training Based Unsupervised Cross-Modality Domain Adaptation for Vestibular Schwannoma and Cochlea Segmentation

Hyungseob Shin[1][0000-0001-7936-5165], Hyeongyu Kim[1][0000-0001-9195-4149], Sewon Kim[0000-0002-3893-252X], Yohan Jun[0000-0003-4787-4760], Taejoon Eo[0000-0002-3546-0184], and Dosik Hwang[0000-0002-2217-2837]

School of Electrical and Electronic Engineering, Yonsei University, Seoul, Korea
whatzupsup@yonsei.ac.kr

**Abstract.** With the advances of deep learning, many medical image segmentation studies achieve human-level performance when in fully supervised condition. However, it is extremely expensive to acquire annotation on every data in medical fields, especially on magnetic resonance images (MRI) that comprise many different contrasts. Unsupervised methods can alleviate this problem; however, the performance drop is inevitable compared to fully supervised methods. In this work, we propose a self-training based unsupervised-learning framework that performs automatic segmentation of Vestibular Schwannoma (VS) and cochlea on high-resolution T2 scans. Our method consists of 4 main stages: 1) VS-preserving contrast conversion from contrast-enhanced T1 scan to high-resolution T2 scan, 2) training segmentation on generated T2 scans with annotations on T1 scans, and 3) Inferring pseudo-labels on non-annotated real T2 scans, and 4) boosting the generalizability of VS and cochlea segmentation by training with combined data (i.e., real T2 scans with pseudo-labels and generated T2 scans with true annotations). Our method showed mean Dice score and Average Symmetric Surface Distance (ASSD) of $0.8570 \pm 0.0705$ and $0.4970 \pm 0.3391$ for VS, $0.8446 \pm 0.0211$ and $0.1513 \pm 0.0314$ for Cochlea on CrossMoDA2021 challenge validation phase leaderboard, outperforming most other approaches.

## 1 Introduction

Domain adaptation (DA) aims to make a model learned from source domain data work well in the target domain without the necessity to perform supervised learning in the target domain. In this paper, we propose a novel unsupervised cross-modality medical image segmentation method and validate it on Vestibular Schwannoma (VS) and cochlea segmentation. Fast and accurate diagnosis of VS is very important in clinical workflows, and the most typical MR protocols used for this include contrast-enhanced T1-weighted (ceT1) and high-resolution T2-weighted (hrT2) scans. To this end, a study of segmenting VS and cochlea from a pair of ceT1 and hrT2 images was presented [1]. However, due to the potential side effect of Gadolinium and long scan time, reducing the cost by segmenting the VS from high-resolution T2 scan alone is being studied [2, 3]. In this study, we aim to perform VS and cochlea segmentation on hrT2 scans using self-training based unsupervised approach trained on unpaired annotated ceT1 and non-annotated hrT2 scans [4, 5]. Our method consists of 4 main stages (Fig. 1): 1) VS-preserving domain translation from contrast-enhanced T1 scan to high-resolution T2 scan followed by self-training scheme which consists of 2) training VS segmentation on generated (i.e., synthetic) T2 scans with annotations on T1 scans, 3) inferring pseudo-labels on non-annotated real T2 scans, and 4) boosting the generalizability of VS and cochlea segmentation by training with combined data (i.e., real T2 scans with pseudo-labels and generated T2 scans with true annotations).

---

[1] H.S and H.K equally contributed to this work.

## 2 Methods

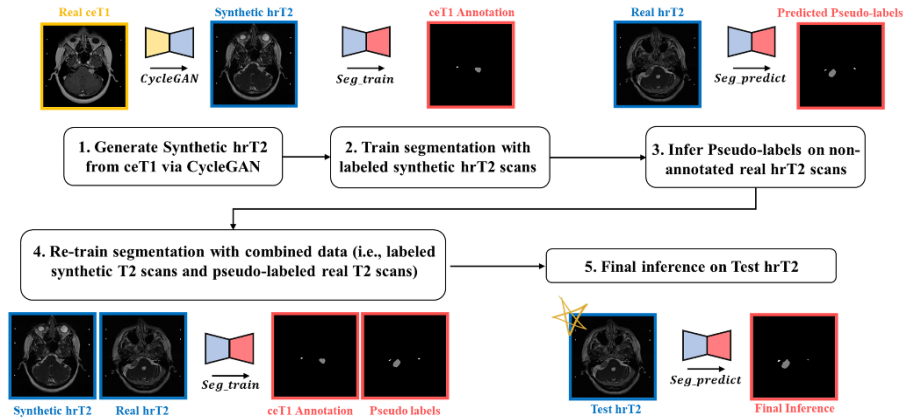

**Fig. 1. Overview of our unsupervised domain adaptation segmentation framework.**

### 2.1 Contrast Conversion from ceT1 to hrT2 scans

We have unpaired datasets of two different MRI contrasts, which are annotated ceT1 MRI scans and non-annotated hrT2 scans. Considering that our task is to predict VS and Cochlea on hrT2 images, we first converted our annotated ceT1 images to hrT2 images using cycleGAN [6], thereby to train segmentation model with the generated hrT2 images afterwards. The proposed contrast conversion network follows the 2D CycleGAN structure, as two 3-level U-net based generators try to convert domains and two PatchGAN [6] discriminators try to distinguish real scans from generated scans in each domain. Based on this structure, we have attached segmentation networks that share the encoder of the generators and predict the VS and cochlea masks from input images (Fig. 2). The segmenters are only trained when the real ceT1 and fake hrT2 scans are fed to the generators since they are paired with the provided annotations. It was observed that the supervision coming from the segmenters helped the shape of VS and cochlea remain well-preserved in the converted images because the segmentation loss made the encoders focus more on areas related to VS (Fig. 4). We also added cycle-consistency loss between the k-space of recovered image and original image. The weighting of adversarial loss, cycle-consistency loss, identity loss and segmentation loss was decided as 1:10:5:100. We used min-max normalization for input images, and the model was trained for 50 epochs using Adam optimizer with an initial learning rate of 0.0001 and multiplicative decay. Segmentation loss on fake hrT2 scan was backpropagated after 5 epochs for stable training.

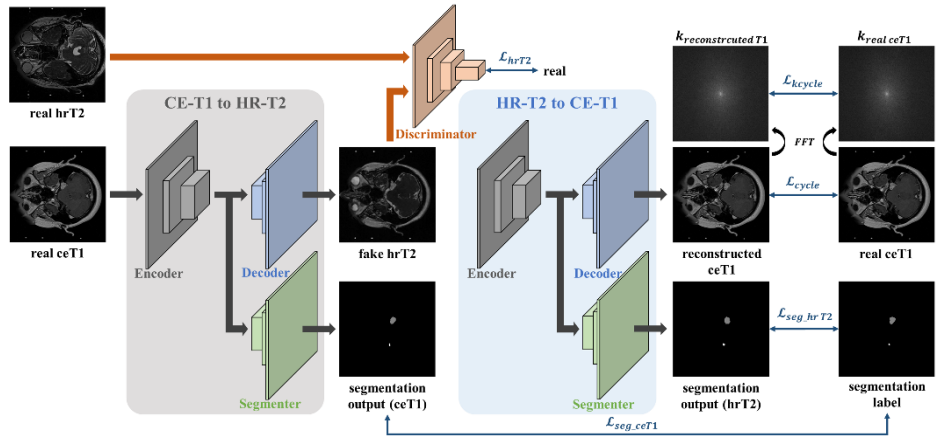

**Fig. 2. Proposed architecture of CycleGAN with additional Segmenters attached to the Encoders of the generators. The reverse direction (real hrT2 to fake ceT1) is omitted for ease of illustration.**

### 2.2 Training segmentation on generated T2 scans and provided annotations

With the generated T2 images and their corresponding segmentation annotations from ceT1 pairs, we can train a segmentation network that segments VS and cochlea segmentation from T2 scans. We employed nnU-Net [7], a standardized and out-of-the-box segmentation framework that self-configures the pre-processing, network architecture, training pipeline from scratch for a given task. The configuration of our 3D U-Net is shown in Fig. 3. It includes deep supervision scheme, input patch size of $40 \times 256 \times 192$, a total of 6 downsampling operations, and initial number of 32 kernels. Each convolutional block consists of convolutional layers followed by instance normalization (IN) and Leaky-ReLU. Dice and cross-entropy loss were used as objective function and the network was trained with stochastic gradient descent with an initial learning rate of 0.01 for 200 epochs. Our motivation was that since the CycleGAN was trained to close the distribution gap between generated hrT2 scans and real hrT2 scans, we could train a segmentation network on synthetic T2 scans with the provided annotations and perform segmentation on real hrT2 scans.

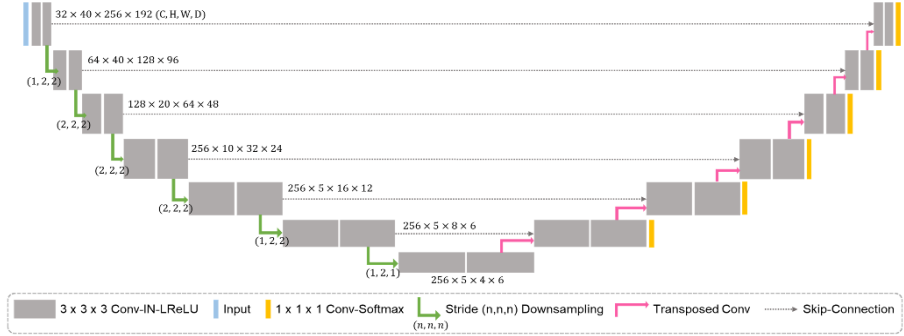

**Fig. 3.** Configured nnU-Net architecture.

### 2.3 Inferring pseudo-labels on Real hrT2 scans to train with combined data

There may still exist some differences in the distribution between generated T2 scans and real T2 scans and this can lead to segmentation performance drop on real T2 scans. To remedy this and boost the generalizability of segmentation on real hrT2 scans, we employed a self-training strategy that additionally uses unlabeled data by utilizing segmentation predictions of real hrT2 scans as pseudo-labels. Therefore, we finally trained our segmentation network in an unsupervised way on combined dataset that consists of generated T2 scans paired with true ceT1 annotations and real T2 scans paired with pseudo-labels. Despite not being real ground truth labels, it has been reported in the previous literature that such self-training scheme also increases the performance and generalizability of the model [8]. For training using combined datasets, 2D nnU-Net (which showed similar architecture to Fig. 3 except dimensionality) and 3D nnU-Net were cross-validated. 2 folds from 2D model and 3 folds from 3D model which showed good performance on validation data were ensembled for final predictions.

## 3 Results

### 3.1 Adding Segmenters on CycleGAN generators

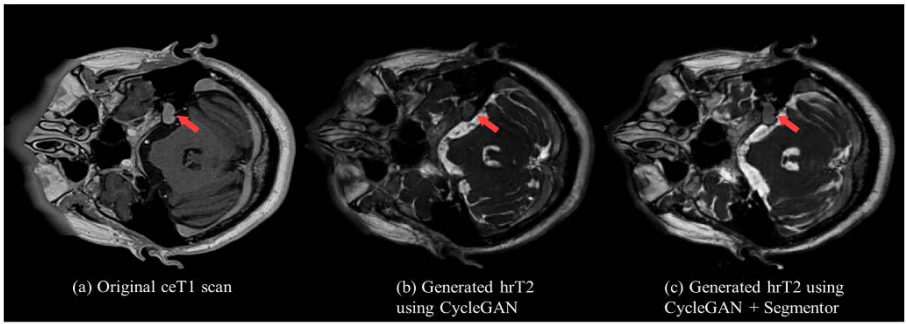

(a) Original ceT1 scan     (b) Generated hrT2 using CycleGAN     (c) Generated hrT2 using CycleGAN + Segmentor

**Fig. 4. Comparison of the generated hrT2 scans between normal CycleGAN and our proposed VS shape-preserving CycleGAN attached with Segmentors.**

From Fig. 4, it is noted that the loss signal from Segmenter clearly contributed to preserving the shape of VS on converted images by making the shared encoder focus on segmentation area and decode the features correspondingly.

### 3.2    Boosting segmentation on real hrT2 using self-training approach

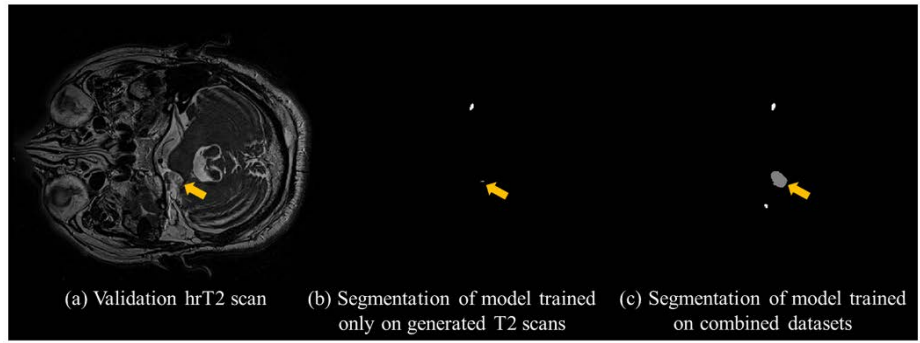

(a) Validation hrT2 scan      (b) Segmentation of model trained      (c) Segmentation of model trained
                                   only on generated T2 scans               on combined datasets

**Fig. 5. Comparison of the segmentation results between the segmentation model trained only on 105 generated T2 scans paired with ceT1 annotations, and the segmentation model trained on 203 combined data (i.e., real T2 scans with pseudo-labels and generated T2 scans with true annotations).**

When the training set for segmentation network is confined to generated T2 scans and the provided annotations on ceT1 scans, the distribution gap between fake hrT2 scans and real hrT2 scans can degrade the segmentation performance on real hrT2 scans. Fig. 5 shows that our self-training approach clearly improved the segmentation performance on validation datasets, particularly those showing different characteristic from generated T2 scans. With self-training scheme, the mean Dice score has quantitatively increased from $0.8088 \pm 0.0822$ to $0.8557 \pm 0.0702$ on the validation phase leaderboard.

## 4    Conclusion

This paper combines contrast conversion and self-training technique for unsupervised cross-modality domain adaptation segmentation. Our method showed mean Dice score and ASSD of $0.8570 \pm 0.0705$ and $0.4970 \pm 0.3391$ for VS, $0.8446 \pm 0.0211$ and $0.1513 \pm 0.0314$ for Cochlea on CrossMoDA2021 challenge validation phase leaderboard, outperforming most other approaches.

## Acknowledgements

This research was supported by Basic Science Research Program through the National Research Foundation of Korea (NRF) funded by the Ministry of Science and ICT (2019R1A2B5B01070488, 2021R1A4A1031437), Brain Research Program through the NRF funded by the Ministry of Science, ICT & Future Planning (2018M3C7A1024734), Y-BASE R&E Institute a Brain Korea 21, Yonsei University, and the Artificial Intelligence Graduate School Program, No. 2020-0-01361, Yonsei University.